# Simultaneous Multi-Slice Diffusion Imaging using Navigator-free Multishot Spiral Acquisition

Yuancheng Jiang [1*], Guangqi Li [1*], Xin Shao [1], and Hua Guo [1#]

[1] Center for Biomedical Imaging Research, School of Biomedical Engineering, Tsinghua University, Beijing, China

* Yuancheng Jiang and Guangqi Li contributed equally to this work.

**Running title:** Navigator-free Multiband Spiral for DWI
**Word Count:** ~4500

# Correspondence:

Hua Guo, PhD

Center for Biomedical Imaging Research

School of Biomedical Engineering

Tsinghua University, Beijing, China

Phone: +86-10-6279-5886

Email: huaguo@tsinghua.edu.cn

# Abstract

**Purpose:** This work aims to raise a novel design for navigator-free multiband (MB) multishot uniform-density spiral (UDS) acquisition and reconstruction, and to demonstrate its utility for high-efficiency, high-resolution diffusion imaging.

**Theory and Methods:** Our design focuses on the acquisition and reconstruction of navigator-free MB multishot UDS diffusion imaging. For acquisition, radiofrequency (RF) pulse encoding was employed to achieve Controlled Aliasing in Parallel Imaging (CAIPI) in MB imaging. For reconstruction, a new algorithm named slice-POCS-enhanced Inherent Correction of phase Errors (slice-POCS-ICE) was proposed to simultaneously estimate diffusion-weighted images and inter-shot phase variations for each slice. The efficacy of the proposed methods was evaluated in both numerical simulation and in vivo experiments.

**Results:** In both numerical simulation and in vivo experiments, slice-POCS-ICE estimated phase variations more precisely and provided results with better image quality than other methods. The inter-shot phase variations and MB slice aliasing artifacts were simultaneously resolved using the proposed slice-POCS-ICE algorithm.

**Conclusion:** The proposed navigator-free MB multishot UDS acquisition and reconstruction method is an effective solution for high-efficiency, high-resolution diffusion imaging.

# 1 Introduction

Diffusion-weighted imaging (DWI) has been widely used in clinical diagnosis and neuroscience research.[1-3] Currently, DWI acquisition primarily relies on single-shot echo-planar imaging (ss-EPI) owing to its rapid imaging speed and resistance to bulk motion. Besides ss-EPI, single-shot spiral (ss-spiral) has emerged as an alternative for DWI acquisition.[4-10] Spiral acquisition offers better sampling efficiency[11] and signal-to-noise ratio (SNR) [4,10,12] over EPI. However, it is also susceptible to B0-field inhomogeneity-induced off-resonance effects, especially when a long spiral readout duration is used for high-resolution imaging.[5] To mitigate this issue, ss-spiral is usually combined with parallel imaging techniques[13,14] to reduce readout duration. However, in parallel imaging, the SNR of images decreases and the aliasing artifacts aggravate with increasing acceleration factors. Therefore, the acceleration factors for ss-spiral are typically restricted and signal averaging is often needed to enhance SNR.

Besides ss-spiral, multi-shot spiral (ms-spiral)[15-18] provides another solution for high-resolution, high-SNR diffusion imaging. A major problem of ms-spiral DWI is the inter-shot phase variations caused by physiological motion (such as respiration, heartbeat, and cerebrospinal fluid pulsation) during diffusion gradient encoding.[19,20] For ms-spiral DWI reconstruction, it is crucial to correct the phase variations for each shot. These phase variations can be measured from extra navigators[21] or self-navigators[15,16], or can be calculated without navigators (navigator-free strategy)[22-25]. Depending on the approach to obtaining the phase information, different spiral acquisition strategies can be divided into various categories.[26-28] Recent studies have demonstrated that navigator-free multishot uniform-density-spiral (UDS) is an efficient acquisition strategy for high-SNR, high-resolution 2D diffusion imaging.[17] Despite that, the acquisition efficiency of multishot UDS diffusion imaging is still not optimal and can be improved.

To improve acquisition efficiency, one commonly employed method is to use multiband (MB) imaging.[29-33] In MB imaging, multiple slices can be excited and acquired simultaneously, and controlled aliasing in parallel imaging (CAIPI) techniques[30,34] are usually used to improve slice separation and minimize g-factor penalties. Presently, there are two main approaches to achieving CAIPI phase encoding: gradient encoding and radiofrequency (RF) pulse encoding.

Gradient encoding usually uses slice gradient ($G_z$) blips to introduce the CAIPI phase[30]. It can be applied to single-shot acquisitions, such as ss-EPI and turbo-spin-echo (TSE), and complicated CAIPI patterns can be realized through flexible gradient design[35]. Several previous studies have also employed this method for MB spiral acquisition in DWI.[36-38] However, this method can also cause some problems, such as through-slice dephasing[29] and the introduction of extra phase for off-isocenter slices[30]. As an alternative, RF encoding imposes the CAIPI phase by modifying RF pulses.[34,39] It can circumvent the aforementioned problems caused by gradient encoding, but is only applicable to multishot acquisitions. To our knowledge, no previous studies have employed RF encoding-based CAIPI for MB spiral acquisitions in DWI. The feasibility and performance of this method in ms-spiral DWI needs investigation.

When using navigator-free MB ms-spiral acquisitions for diffusion imaging, the image reconstruction process becomes inherently challenging because both inter-shot phase variations and slice aliasing need to be considered. Although various methods (SENSE+CG[23], POCS-ICE[25], etc.[22,24,40,41]) have been proposed to reconstruct navigator-free single-band (SB) ms-spiral DWI data, these methods are not suitable for MB DWI reconstruction because of slice aliasing. One way to solve this problem is to first use either non-Cartesian split-slice-GRAPPA (NCSG)[42] or direct-spiral slice-GRAPPA (ds-SG)[43] to resolve slice aliasing, and then apply the aforementioned SB reconstruction techniques to resolve inter-shot phase variations slice by slice. However, the effectiveness of this two-step method may be compromised due to the concurrent presence of inter-shot phase variations and slice aliasing. Another approach involves leveraging the 3D k-space framework of MB imaging[44-46] and simultaneously reconstructing multiple slices using model-based methods[36-38]. Nevertheless, the implementation of these methods is also challenging due to inter-shot phase variations.

In this study, we aimed to propose a novel design for navigator-free MB multishot UDS acquisition and reconstruction, and to utilize it for high-efficiency, high-resolution diffusion imaging. During acquisition, RF encoding was used to accomplish CAIPI for MB sampling. For reconstruction, a new method called slice-POCS-enhanced Inherent Correction of phase Errors (slice-POCS-ICE) was proposed to simultaneously estimate multi-slice DWI images and inter-shot phase variations. Both numerical simulation and in vivo experiments were conducted to evaluate the performance of our proposed methods.

## 2 Theory

### 2.1 RF Encoding for CAIPI

To generate MB RF pulses, one simple way is to sum multiple frequency-modulated SB RF pulses[29]:

$$RF_{MB}(t) = \sum_{n=1}^{N_{MB}} RF(t)e^{j\omega_n t} \quad [1]$$

where $RF_{MB}(t)$ is the MB RF pulse that simultaneously excites multiple slices, $RF(t)$ is a SB RF pulse, $N_{MB}$ is the number of simultaneously excited slices and $\omega_n$ is the center frequency of each slice.

In this study, RF encoding-based CAIPI is employed. Specifically, the MB RF pulse for each shot is:

$$RF_{MB,i}(t) = \sum_{n=1}^{N_{MB}} RF(t)e^{j\omega_n t}\phi_{n,i}, \quad i = 1 \ldots N_{shot} \quad [2]$$

where $RF_{MB,i}(t)$ is the MB RF pulse of shot $i$, $N_{shot}$ is the number of shots, and $\phi_{n,i}$ denotes the CAIPI phase information introduced by shot $i$ for slice $n$, as expressed by $\phi_{n,i} = e^{-j2\pi(n-1)(i-1)/N_{MB}}$. $\phi_{n,i}$ is different for each shot and each slice.

With the CAIPI phase modulation introduced here, the acquisition of MB ms-spiral data can be regarded as sampling in a 3D k-space[44,45]. This 3D k-space is formed by the $k_x$-$k_y$ plane and an additional dimension, $k_{mb}$. If this k-space is fully sampled, a 3D volume can be reconstructed using 3D non-uniform Fourier transform (NUFFT)[47]. In this work, this 3D k-space is undersampled along $k_{mb}$ and needs to be recovered during reconstruction. For clarity, Figure 1A illustrates the 3D k-space formed by the proposed MB ms-spiral acquisition.

### 2.2 slice-POCS-ICE

In the data acquisition process of multishot DWI, the phase of different shots varies due to physiological motion during diffusion gradient encoding. Therefore, when the RF pulse in Eq. 2 is utilized for MB ms-spiral DWI acquisition, the acquired spiral data of $i$-th shot and $j$-th coil should be expressed as,

$$d_{i,j} = \sum_{n=1}^{N_{MB}} \phi_{n,i} G_i \mathcal{F} S_{n,j} \psi_{n,i} I_n \qquad [3]$$

where $I_n$ is the DWI image to be reconstructed, $S_{n,j}$ denotes the coil sensitivity maps of slice $n$, coil $j$, $\mathcal{F}$ is the discrete Fourier transform (DFT) operation and $G_i$ is the operation that transfers data from Cartesian grid to the sampling location of $i$-th shot. $\psi_{n,i}$ is the phase variation for slice $n$, shot $i$, and $\phi_{n,i}$ is the introduced CAIPI phase. After $N_{shot}$ excitations and acquisitions, MB ms-spiral data can be collected. For clarity, a schematic diagram that illustrates the formation process of $d_{i,j}$ and the MB data for the [$N_{shot} = 2, N_{MB} = 2$] case is shown in Figure 1B.

The image reconstruction process aims to estimate $I_n$ from Eq. 3. In Eq. 3, $d_{i,j}$, $\phi_{n,i}$, and $S_{n,j}$ are known variables, while the phase variation $\psi_{n,i}$ is unknown and needs to be estimated. For SB ms-spiral DWI, various methods[22-25,40,41] have been proposed to calculate the inter-shot phase variations and reconstruct DWI images. However, these methods are not suitable for MB data because the undersampled data along the slice direction needs to be recovered during reconstruction. Here, we proposed a new algorithm specialized for MB ms-spiral reconstruction, and we refer to it as slice-POCS-enhanced Inherent Correction of phase Errors (slice-POCS-ICE).

Slice-POCS-ICE inherits the core concept of POCS-ICE[25], which is to estimate the phase variations and DWI images simultaneously during reconstruction. Hence, the reconstruction process of slice-POCS-ICE can be described as an optimization problem:

$$\min_{\psi_{n,i}, I_n} \sum_{i=1}^{N_{shot}} \sum_{j=1}^{N_{coil}} \left\| \sum_{n=1}^{N_{MB}} \phi_{n,i} G_i \mathcal{F} S_{n,j} \psi_{n,i} I_n - d_{i,j} \right\|_2^2 \qquad [4]$$

where $N_{shot}$ and $N_{coil}$ denote the number of shots and receive coils, respectively.

Slice-POCS-ICE employs the projecting onto convex set (POCS) algorithm[48] to solve Eq. 4. POCS addresses reconstruction problems by iteratively projecting data onto a predefined convex set $\Omega$. In slice-POCS-ICE, we designed a tailored data projection operation to reconstruct the MB ms-spiral DWI data, described as follows.

*2.2.1 Tailored Data Projection Operation*

For POCS-based reconstruction, Ω is usually a convex set that consists of images whose k-space values are identical to the acquired data at sampled locations.[49] Moreover, for ms-spiral diffusion imaging, due to the presence of inter-shot phase variations, the data projection is applied to each shot and each coil separately. Specifically, the data projection operation for shot $i$, coil $j$ in SB ms-spiral DWI can be defined as,

$$I_{i,j}^{\Omega} = I_{i,j} + \mathcal{F}^{-1}G_i^{-1}(d_{i,j} - G_i\mathcal{F}I_{i,j}) \quad [5]$$

where $I_{i,j}$ and $d_{i,j}$ are the estimated image and the acquired k-space data of shot $i$, coil $j$, respectively. $I_{i,j}^{\Omega}$ is the image after data projection. $\mathcal{F}$ and $\mathcal{F}^{-1}$ are DFT and inversed DFT (IDFT). $G_i$ is the operation that transfers data from the Cartesian grid to the sampling location of $i$-th shot, while $G_i^{-1}$ denotes the opposite operation (transform from spiral trajectory to Cartesian grid). For SB imaging, the images after data projection will fall into the convex set Ω, and iteratively applying this operation will lead to convergence and finish the reconstruction process.

The data projection operation in Eq. 5 cannot lend itself to MB imaging because the MB data contain aliasing signals along slice direction. To address this issue, the 3D k-space framework introduced in section 2.1 is employed here and a tailored data projection operation for MB data is proposed. In short, this data projection operation takes the simultaneously excited 2D images as input, and outputs their new estimations. Figure 2 depicts its diagram and the detailed steps are described as follows.

First, the input simultaneously excited 2D images are used to construct a 3D image. For shot $i$, coil $j$, supposing the input images are $\{I_{1,i,j}, ..., I_{N_{MB},i,j}\}$, the 3D image $U_{i,j}$ is constructed by directly stacking $I_{1,i,j}$ to $I_{N_{MB},i,j}$ along the slice dimension.

Second, the constructed 3D image $U_{i,j}$ is fed to a 3D data projection operation, defined as,

$$U_{i,j}^{\Omega} = U_{i,j} + \mathcal{F}_{3D}^{-1}G_{3D,i}^{-1}(D_{i,j} - G_{3D,i}\mathcal{F}_{3D}U_{i,j}) \quad [6]$$

where $U_{i,j}^{\Omega}$ represents the 3D image after data projection. $D_{i,j}$ is the acquired 3D k-space data, which is formed by placing $d_{i,j}$ (the acquired MB k-space data) at the corresponding position in 3D k-space. $\mathcal{F}_{3D}$ and $\mathcal{F}_{3D}^{-1}$ are 3D DFT and 3D IDFT operators. $G_{3D,i}$ is the operation to transfer data from the 3D Cartesian grid to the same location as $d_{i,j}$ in the 3D k-space and $G_i^{-1}$ denotes the opposite operation.

Third, the 3D image after data projection ($U_{i,j}^{\Omega}$) is split into multiple 2D images to yield the new estimations for input images $\{I_{1,i,j}^{\Omega}, \dots, I_{N_{MB},i,j}^{\Omega}\}$.

After the aforementioned three steps, the simultaneously excited 2D images undergo concurrent data projection, resulting in new estimations. These three steps can be integrated into a single operation, denoted as $P_{ij}$, which is a data projection operation tailored for MB ms-spiral reconstruction and can be expressed as follows,

$$\{I_{1,i,j}^{\Omega}, \dots, I_{N_{MB},i,j}^{\Omega}\} = P_{ij}\{I_{1,i,j}, \dots, I_{N_{MB},i,j}\} \qquad [7]$$

*2.2.2 Algorithm*

Slice-POCS-ICE is an iterative method containing three consecutive steps in each iteration: phase variation recovery, data update, and shot averaging. Figure 3 shows the schematic diagram of slice-POCS-ICE, and the detailed steps of each iteration are explained here. In the following illustration, the upper right corner marker indicates the number of iterations. For example, $I_n^{(m)}$ represents the estimated image of the $n$-th slice, after $m$ iterations.

Step 1, phase variation recovery. The estimated phase variations $\psi_{n,i}^{(m)}$ are added to the estimated images $I_n^{(m)}$ to generate multishot images:

$$I_{n,i}^{(m)} = I_n^{(m)} \cdot \psi_{n,i}^{(m)} \qquad [8]$$

where $I_{n,i}$ denotes the estimated image of slice $n$ and shot $i$.

Step 2, data update. In this step, the multishot images are updated by data projection.

First, the single-coil multishot images are expanded to multi-coil multishot images:

$$I_{n,i,j}^{(m)} = S_{n,j} \cdot I_{n,i}^{(m)} \qquad [9]$$

where $I_{n,i,j}$ is the estimated image of slice $n$, shot $i$, and coil $j$ and $S_{n,j}$ denotes the coil sensitivity maps of slice $n$, coil $j$. The multi-coil images then undergo the data projection operation $P_{ij}$ for new estimation:

$$\{I_{1,i,j}^{(m+1)}, \dots, I_{N_{MB},i,j}^{(m+1)}\} = P_{ij}\{I_{1,i,j}^{(m)}, \dots, I_{N_{MB},i,j}^{(m)}\} \qquad [10]$$

After data projection, the multi-coil images are coil-combined to yield the updated multishot images:

$$I_{n,i}^{(m+1)} = \sum_{j=1}^{N_{coil}} \frac{S_{n,j}^* I_{n,i,j}^{(m+1)}}{\sum_{k=1}^{N_{coil}} S_{n,k}^* S_{n,k}} \quad [11]$$

where * means complex conjugate.

Step 3, shot averaging. In this step, new DWI images and phase variations are estimated.

The new phase variations are extracted from the updated multishot images, and a low-pass filter is used to filter them. Specifically, k-space filtering is utilized, i.e.,

$$\tilde{I}_{n,i} = \text{IDFT}\left(W \cdot \text{DFT}\left(I_{n,i}^{(m+1)}\right)\right) \quad [12]$$

$$\psi_{n,i}^{(m+1)} = \tilde{I}_{n,i}/|\tilde{I}_{n,i}| \quad [13]$$

where the $\text{DFT}(\cdot)$ denotes DFT operation and $\text{IDFT}(\cdot)$ denotes IDFT operation. $W$ is a 2D triangular window applied in k-space and $|\cdot|$ denotes the operation to get the magnitude of an image. After Eq. 12 and 13, the new phase variations are estimated.

The new estimated DWI images are generated by averaging all the shots after removing the phase variations:

$$I_n^{(m+1)} = \frac{1}{N} \sum_{i=1}^{N} \left[\psi_{n,i}^{(m+1)}\right]^* I_{n,i}^{(m+1)} \quad [14]$$

The aforementioned three steps will be repeated until the algorithm reaches convergence. Slice-POCS-ICE has two convergence conditions. One is that the overall algorithm update, denoted by $u_{all}^{(m+1)}$, is smaller than a predefined tolerance $\tau$. To get $u_{all}^{(m+1)}$, the update of each slice is first calculated as,

$$u_n^{(m+1)} = \frac{\left\|I_n^{(m+1)} - I_n^{(m)}\right\|_F}{\left\|I_n^{(m)}\right\|_F}, \quad n = 1 \ldots N_{MB} \quad [15]$$

where $u_n^{(m+1)}$ is the algorithm update for slice $n$ after $m+1$ iterations and $\|\cdot\|_F$ is the Frobenius norm. The overall update is defined as,

$$u_{all}^{(m+1)} = \max\left(u_1^{(m+1)}, \ldots, u_{N_{MB}}^{(m+1)}\right) \quad [16]$$

The $\max(\cdot)$ function outputs the maximum value among the input numbers, ensuring that all slices have converged when $u_{all}^{(m+1)} < \tau$. Another convergence condition is met when a predefined maximum iteration number is reached. The algorithm terminates immediately when either of these two convergence conditions is satisfied. Otherwise, it proceeds to the next iteration.

## 3 Methods

All MR data were acquired from three healthy volunteers on a Philips 3.0T Ingenia CX scanner (Philips Healthcare, Best, The Netherlands) with a 32-channel head coil. A consent form approved by the local Institutional Review Board at Tsinghua University was obtained from every volunteer before scan.

### 3.1 Numerical simulation

To verify our proposed acquisition and reconstruction methods, numerical simulation was performed on a T2-weighted multi-coil head volume from a healthy volunteer. To begin with, suppose the shot number was $N_{shot}$ and the number of simultaneously excited slices was $N_{MB}$, the procedure to create the simulated MB ms-spiral data was as follows.

(a) Select $N_{MB}$ slices from the head volume. Use ESPIRiT[50] to calculate the sensitivity maps for each slice. (b) Coil-combine the selected data to obtain the reference images $I_{ref}$. (c) Generate $N_{shot} \times N_{MB}$ spatially-varying second-order phase maps randomly to simulate the phase variations for each shot and slice.[41] (d) Multiply $I_{ref}$ by the phase maps and the sensitivity maps to obtain multishot multi-coil data. (e) Transform the multi-coil image to k-space spiral data using NUFFT. (f) Perform CAIPI phase modulation on each slice, and directly sum multiple slices to generate MB k-space data.

The simulation data was then reconstructed by different methods, and the reconstruction quality was evaluated by the normalized root mean square error (nRMSE), which is defined as:

$$\text{nRMSE} = \frac{\|I - I_{ref}\|_F}{\|I_{ref}\|_F} \qquad [17]$$

where $I$ is the reconstruction result.

## 3.2 in vivo experiments

Three in vivo experiments were conducted to validate the proposed acquisition and reconstruction methods. Diffusion-weighted data were acquired using the Stejskal-Tanner spin-echo sequence[51], with the proposed navigator-free MB multishot UDS trajectory for readout. The MB pulses were generated by summing $N_{MB}$ phase modulated SB sinc pulses. In addition, second-order shimming was used to reduce B0 inhomogeneity, and low-resolution B0 field maps were acquired by a 2D multi-echo gradient echo sequence for off-resonance correction. For all data acquired, the detailed scanning parameters are summarized in Table 1.

In the first experiment, a basic in vivo validation was conducted. Diffusion-weighted data were collected using MB and SB acquisitions, and the common parameters for different acquisitions were: FOV = 210 × 210 mm$^2$, b = 800 s/mm$^2$, TE/TR = 60/3000 ms. In the second experiment, a higher shot number of 6 was used to acquire high-resolution images, with resolution = 1×1×3 mm$^3$ and $N_{MB}$ = 2. In the third experiment, diffusion images with 1.5 mm isotropic resolution and whole-brain coverage were acquired to further evaluate our method. Parameters include: $N_{shot}$ = 3, $N_{MB}$ = 2, number of signal averages (NSA) = 3.

For each MB scan in each experiment, a calibration scan was also conducted. The calibration scan shared the same parameters as the MB scan, except that it used SB acquisitions and did not collect DWI data.

## 3.3 Data processing

Slice-POCS-ICE and the two-step methods were employed for reconstruction after acquisition, and their respective performances were compared. For the two-step methods, the first step is to recover aliasing signals of MB data using NCSG, and the second step is to reconstruct each slice using SB reconstruction algorithms (POCS-ICE or SENSE+CG).

In slice-POCS-ICE, the initial images were set to zero. The sensitivity maps for the in vivo experiments were calculated from the SB calibration data. The width of the triangular window $W$ in step 3 was set to half of the matrix size. Additionally, the algorithm tolerance $\tau$ was $10^{-6}$, and the maximum iteration number was 200.

In the two-step methods, NCSG was applied to each shot separately due to inter-shot phase variations. The NCSG kernel was trained on the original SB head volume for numerical simulation, and on the SB calibration data for in vivo experiments, with a kernel size set to 7×7. In the second step, the sensitivity maps for in vivo experiments were calculated from the calibration data. For POCS-ICE, the initial images were set to zero, with a tolerance of $10^{-6}$ and a maximum iteration number of 200. For SENSE+CG, the tolerance was $10^{-6}$, and the maximum iteration number for the CG step was 12.

After reconstruction, conjugate phase correction[52] was employed for static B0 off-resonance correction, with $\Delta f$ information derived from the acquired field maps. In addition, fractional anisotropy (FA) maps were calculated using the FMRIB Software Library (FSL) software[53].

All the image reconstruction and post-processing steps were implemented in Matlab (MathWorks Inc., Natick, MA) on a Linux workstation (CPU: AMD Ryzen, 4.6 GHz, 12 Cores; RAM: 128GB). In the [$N_{shot} = 3$, $N_{MB} = 2$] case, the computation time for slice-POCS-ICE to reconstruct $N_{MB}$ slices was approximately 1 minute, or 30 seconds per slice. For the two-step methods, NCSG took 2 minutes in the first step. After that, POCS-ICE and SENSE+CG took about 50 seconds and 10 seconds to reconstruct one slice. Therefore, the total reconstruction time for one slice is 110 seconds and 70 seconds, respectively.

## 4 Results

### 4.1 Numerical simulation

Figure 4 shows the phase variation reconstruction results for the [$N_{shot} = 3$, $N_{MB} = 2$] case. The results of slice-POCS-ICE are close to the ground truth and contain no visible errors. In contrast, discernible errors can be observed in the results of the two-step methods, regardless of whether the second step employs SENSE+CG or POCS-ICE. As shown in Figure S1, the proposed method was proven effective even in the challenging case of [$N_{shot} = 4$, $N_{MB} = 3$]. Slice-POCS-ICE continues to estimate the phase variations more accurately than the two-step methods, while the estimation error of the two-step methods increases as $N_{shot}$ and/or $N_{MB}$ increases, as evident when comparing Figure 4 and Figure S1.

The effectiveness of the proposed methods was further validated by examining the image reconstruction results in numerical simulation. Figures 5 and S2 show the results of each method for the [$N_{shot}$ = 3, $N_{MB}$ = 2] and [$N_{shot}$ = 4, $N_{MB}$ = 3] cases, respectively, and the nRMSE values between the results and the ground truth are labeled. In both cases, slice-POCS-ICE consistently produces results with higher image quality and fewer artifacts than the two-step methods, as evidenced by the lowest nRMSE (2.4% ± 0.4% in the [$N_{shot}$ = 3, $N_{MB}$ = 2] case, and 8.8% ± 1.2% in the [$N_{shot}$ = 4, $N_{MB}$ = 3] case). Conversely, the results of the two-step methods exhibit higher nRMSE (~18% in the [$N_{shot}$ = 3, $N_{MB}$ = 2] case, and ~30% in the [$N_{shot}$ = 4, $N_{MB}$ = 3] case) and visible reconstruction errors, especially in the [$N_{shot}$ = 4, $N_{MB}$ = 3] case. The inferior performance of the two-step methods may be attributed to their limited ability to estimate inter-shot phase variations, as illustrated in Figure 4 and S1.

**4.2 in vivo experiments**

In the first in vivo experiment, slice-POCS-ICE was first compared with the two-step methods. Figure 6 and S4 show the image reconstruction results of each method for the [$N_{shot}$ = 4, $N_{MB}$ = 2] and [$N_{shot}$ = 4, $N_{MB}$ = 3] cases, respectively. In both cases, slice-POCS-ICE demonstrates good reconstruction quality, exhibiting minimal artifacts with favorable SNR. The results of the two-step methods, however, contain noticeable reconstruction errors, especially in the [$N_{shot}$ = 4, $N_{MB}$ = 3] case. This observation is consistent with the findings in numerical simulation, indicating that in high-$N_{shot}$ and/or high-$N_{MB}$ scenarios, the two-step methods cannot estimate the inter-shot phase variations accurately, resulting in poorer reconstruction performance.

Our method was also compared with SB acquisition and reconstruction methods for further evaluation, and the results are shown in Figures 6 and S5. Because of MB acceleration, the SNR of the MB images is slightly lower than that of the SB images. However, in both $N_{MB}$ = 2 (Figure 6) and $N_{MB}$ = 3 (Figure S5) cases, the MB images exhibit favorable structural consistency with the SB images, in both DWI and color-coded FA (cFA) images. These results demonstrate that high-fidelity diffusion images can be obtained using our proposed MB acquisition and reconstruction method.

Figures 8 and S6 display several representative slices from the skull base to the top, showcasing the reconstruction results of slice-POCS-ICE for the [$N_{shot}$ = 4, $N_{MB}$ = 3] and [$N_{shot}$ = 3, $N_{MB}$ = 2] cases, respectively. The images are free of visible artifacts, and fine brain structures can be observed in the cFA images.

In the second and third in vivo experiments, our method was used to acquire high-resolution diffusion images, as presented in Figures S7 and 9, respectively. Figure S7 shows the images with a resolution of 1×1×3 mm³, obtained from the [$N_{shot}$ = 6, $N_{MB}$ = 2] acquisition. Figure 9 shows the isotropic 1.5 mm resolution images from the [$N_{shot}$ = 3, $N_{MB}$ = 2] acquisition. Due to the increase in image resolution, there is a slight SNR decrease in the results, especially in the cFA images (Figure 9). Nevertheless, the reconstruction images still maintain satisfactory quality without significant artifacts, indicating that our acquisition and reconstruction workflow is effective for high-resolution diffusion imaging.

## 5 Discussion

In this study, we proposed an acquisition and reconstruction workflow for the navigator-free MB multishot UDS sequence and utilized it for diffusion imaging. For DWI data acquisition, MB RF pulses were used to simultaneously excite multiple slices, and RF phase modulation was employed to implement CAIPI. For DWI image reconstruction, we developed a novel algorithm named slice-POCS-ICE, which simultaneously estimates the inter-shot phase variations and diffusion images for each slice. Both numerical simulation and in vivo experiments demonstrated the superior performance of slice-POCS-ICE over other two-step methods (NCSG and SENSE+CG, and NCSG and POCS-ICE). Through the proposed MB acquisition and reconstruction workflow, we successfully achieved high-efficiency, high-resolution spiral-based diffusion imaging.

During the MB data acquisition process, we employed RF encoding, instead of the commonly used gradient encoding, to realize CAIPI. This strategy can mitigate the effects of $G_z$ blip gradients for spiral imaging. First, $G_z$ blips introduce an extra phase to the off-isocenter slices, and this extra phase needs to be removed during reconstruction.[30] Second, $G_z$ blips can cause through-slice dephasing, especially for thick-slice acquisitions.[29] Third, spiral acquisition is sensitive to dynamic field imperfections.[5,54] Issues such as gradient delays[55], eddy currents[56], and concomitant fields[57] usually lead to deviations in the spiral trajectory, resulting in ringing and blurring artifacts in the final images.[5,54,58,59] The introduction of $G_z$ gradients during spiral acquisition may exacerbate these issues, making spiral trajectory correction more challenging. To address this challenge, Engel et al.[37] used field probes to detect dynamic field information for reconstruction, and Wu et al.[38] employed gradient impulse response function (GIRF) for trajectory correction. As an alternative, RF encoding can achieve CAIPI phase modulation without considering the effects of $G_z$ gradients. Moreover, it is straightforward to implement and naturally lends itself to multishot acquisition. Therefore, RF encoding was utilized to achieve CAIPI in this work.

In the reconstruction process, we proposed slice-POCS-ICE, which iteratively corrects motion-induced phase variations and reconstructs navigator-free multishot DWI images. A crucial step in slice-POCS-ICE is the tailored data projection operation, which is designed within a 3D k-space framework and can update multiple slices simultaneously. Both numerical simulation and in vivo experiments demonstrate that slice-POCS-ICE outperforms the two-step methods (NCSG and SENSE+CG, and NCSG and POCS-ICE). For the two-step methods, the NCSG kernel is trained on the SB b = 0 data. However, owing to inter-shot phase variations, the capability of NCSG to resolve MB aliasing for multishot DWI data can be compromised. When residual aliasing artifacts remain in the images, SENSE+CG or POCS-ICE in the second step will also fail to accurately estimate the phase variations, leading to a decline in the final reconstruction quality. In contrast, slice-POCS-ICE employs a model-based strategy to simultaneously resolve inter-shot phase variations and MB aliasing artifacts, independent of the performance of NCSG.

Like POCS-ICE, slice-POCS-ICE demonstrates stable convergence behavior. Figure S3 shows the nRMSE curve for each slice in numerical simulation. For both [$N_{shot}$ = 3, $N_{MB}$ = 2] and [$N_{shot}$ = 4, $N_{MB}$ = 3] cases, the nRMSE for each slice decreases monotonically as the iterations proceed, with similar descending trends observed across different slices. These results suggest that slice-POCS-ICE updates each slice synchronously, gradually improving the estimation accuracy of both images and phase variations as iterations progress. However, slice-POCS-ICE requires a relatively high number of iterations to reach convergence, and this number increases with $N_{MB}$ or $N_{shot}$ (~60 iterations for [$N_{shot}$ = 3, $N_{MB}$ = 2], and ~150 iterations for [$N_{shot}$ = 4, $N_{MB}$ = 3]), resulting in long reconstruction times. Despite this, its reconstruction time is still shorter than that of the two-step methods used in this study (30 s/slice vs. 110 or 70 s/slice). To further reduce the reconstruction time of slice-POCS-ICE, the algorithm can be accelerated using FISTA[40,60] or by employing parallel computing.

Through the proposed acquisition and reconstruction workflow, we have achieved high-efficiency, high-resolution diffusion imaging. This workflow can naturally be extended to other spiral acquisition sequences, such as multishot variable-density spiral (VDS)[28] or dual-density spiral (DDS)[27]. In this study, DWI images with resolutions of 1.5 mm isotropic and 1×1×3 mm$^3$ were acquired using 3-shot and 6-shot UDS (as shown in Figures 9 and S7), respectively. A larger $N_{shot}$ and/or $N_{MB}$ can be used to obtain images with higher resolution. However, when $N_{shot}$ or $N_{MB}$ is too high, slice-POCS-ICE might not be able to reconstruct the images because the net acceleration factor for each shot ($R_{net} = N_{shot} \times N_{MB}$) is too large. Despite that, if a head coil with more receive channels (e.g., 64-channel head coil) could be utilized, the acceleration performance can be further improved.[61]

This study has several limitations. First, the MB RF pulse employed in this study is designed by a simple summation of SB sinc pulses, resulting in a long pulse duration and thereby elongating TE, which reduces SNR. Some pulse design methods such as VERSE[62] or root-flipped[63] can be used to shorten TE and improve SNR. Second, in this study, the eddy currents across different diffusion directions[5,64] were not corrected. To address this issue, field probes can be used to capture dynamic field information[65,66], which can be incorporated into reconstruction for better image quality.[67,68] Third, residual B0-inhomogeneity-related artifacts are still visible in regions like frontal sinuses and ear canals. To reduce this artifact, dynamic field shimming techniques[69-71] can be used to reduce B0 field inhomogeneity, and some advanced deblurring algorithms[72-76] can be employed. Last, although we have enhanced acquisition efficiency by using MB acquisition, the inherent SNR limitation of 2D acquisition prevents us from acquiring higher-resolution images. In the future, we plan to explore 3D spiral acquisition and reconstruction techniques[77-79] to achieve submillimeter-resolution diffusion imaging.

## 6 Conclusions

In this study, we designed a novel method for navigator-free MB multishot UDS acquisition and reconstruction and demonstrated its utility for diffusion imaging. We used RF encoding to accomplish CAIPI for acquisition and proposed slice-POCS-ICE for reconstruction. Through the proposed workflow, we successfully achieved high-efficiency, high-resolution spiral-based diffusion imaging.

## Acknowledgements

This work was supported by the National Key Research and Development Program of China (grant number: 2022YFC2405303).

# Figures

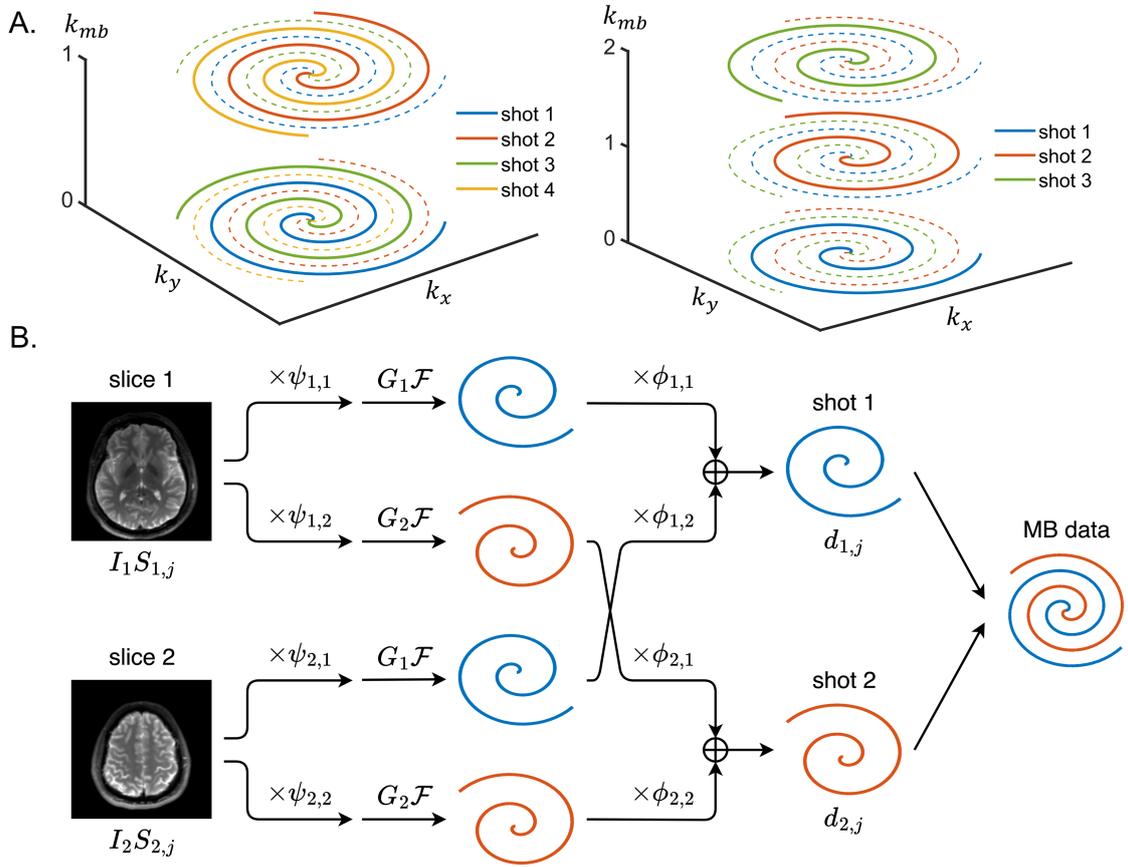

**Figure 1** Diagram of the multiband (MB) multishot spiral acquisition. **(A)** Illustration of the 3D k-space formed by the MB data. The acquired and unacquired data are respectively represented by the solid and dashed lines, and the data from different shots are distinguished by different colors. From left to right: the 3D k-space from the $[N_{shot} = 4, N_{MB} = 2]$ and $[N_{shot} = 3, N_{MB} = 3]$ acquisitions, respectively. **(B)** The formation process of MB diffusion data. For clarity, the process of a single coil $j$ from the $[N_{shot} = 2, N_{MB} = 2]$ case is illustrated. First, the signal from each slice is multiplied by phase variations and transformed into k-space, yielding single-band (SB) k-space spiral data. Next, the SB data are modulated by the CAIPI phase and summed to form the MB data.

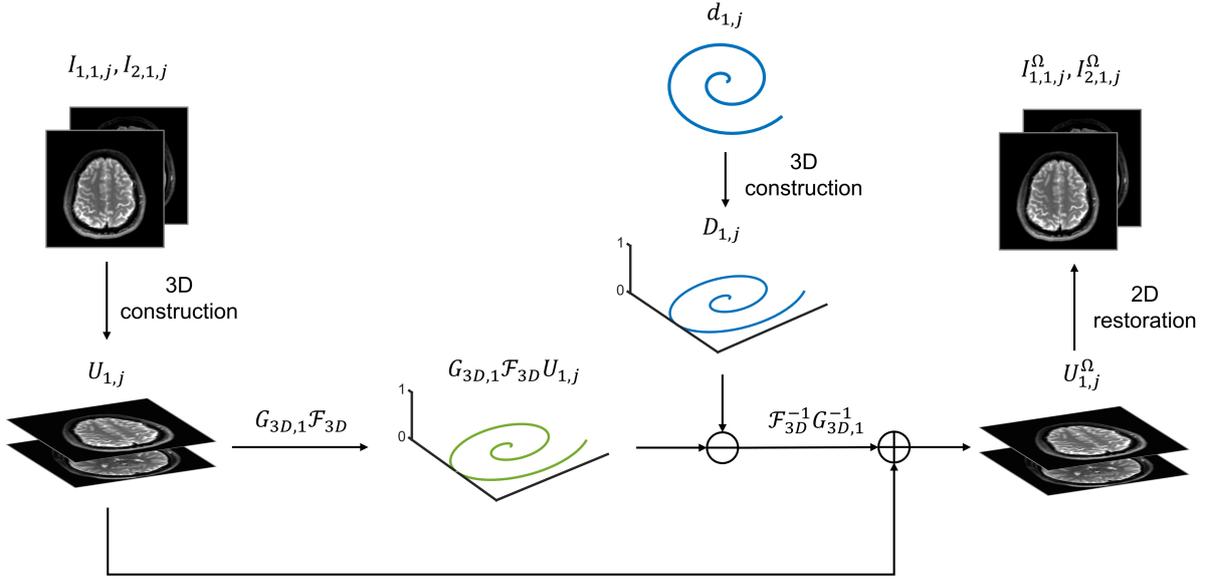

**Figure 2** Diagram of the tailored data projection operation for slice-POCS-ICE. For clarity, the process of shot 1 and coil $j$ from the [$N_{shot} = 2, N_{MB} = 2$] case is shown here. First, the 3D image $U_{1,j}$ is constructed from the input images ($I_{1,1,j}$, $I_{2,1,j}$), and the 3D k-space data $D_{1,j}$ is constructed from the acquired MB data $d_{1,j}$. Next, $U_{1,j}$ is transformed into 3D k-space and subtracted from $D_{1,j}$. The resulting difference ($D_{1,j} - G_{3D,1}\mathcal{F}_{3D}U_{1,j}$) is then transformed back into image domain and added to $U_{1,j}$ to yield the 3D image after data projection ($U_{1,j}^{\Omega}$). Finally, multi-slice 2D images after data projection ($I_{1,1,j}^{\Omega}$, $I_{2,1,j}^{\Omega}$) are restored from $U_{1,j}^{\Omega}$.

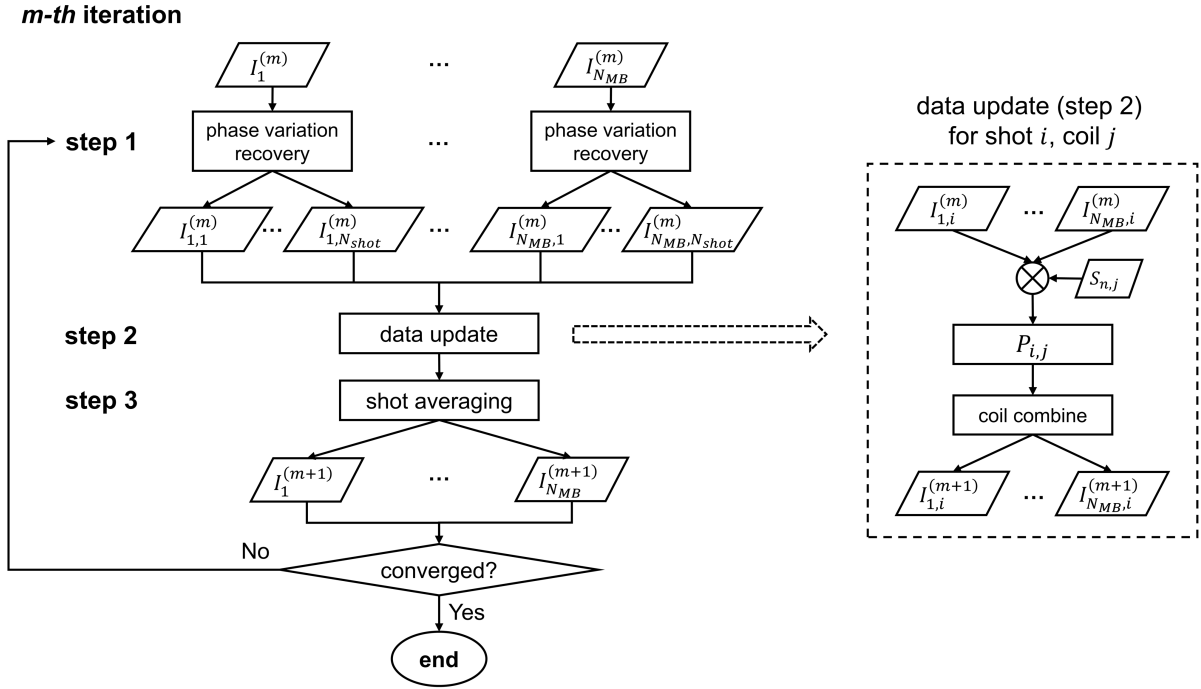

**Figure 3** Diagram of slice-POCS-ICE. The left part shows the flow chart of the algorithm. There are three major steps in each iteration: phase variation recovery, data update, and shot averaging. In step 1, the estimated phase variations are added to the estimated images to generate multishot images; In step 2, the multishot images are updated by data projection; In step 3, new phase variations and images are estimated from the updated multishot images. The right part shows the details of the data update step. First, the single-coil images are expanded to the multi-coil images. After that, the multi-coil images are updated by data projection ($P_{i,j}$) and then coil-combined.

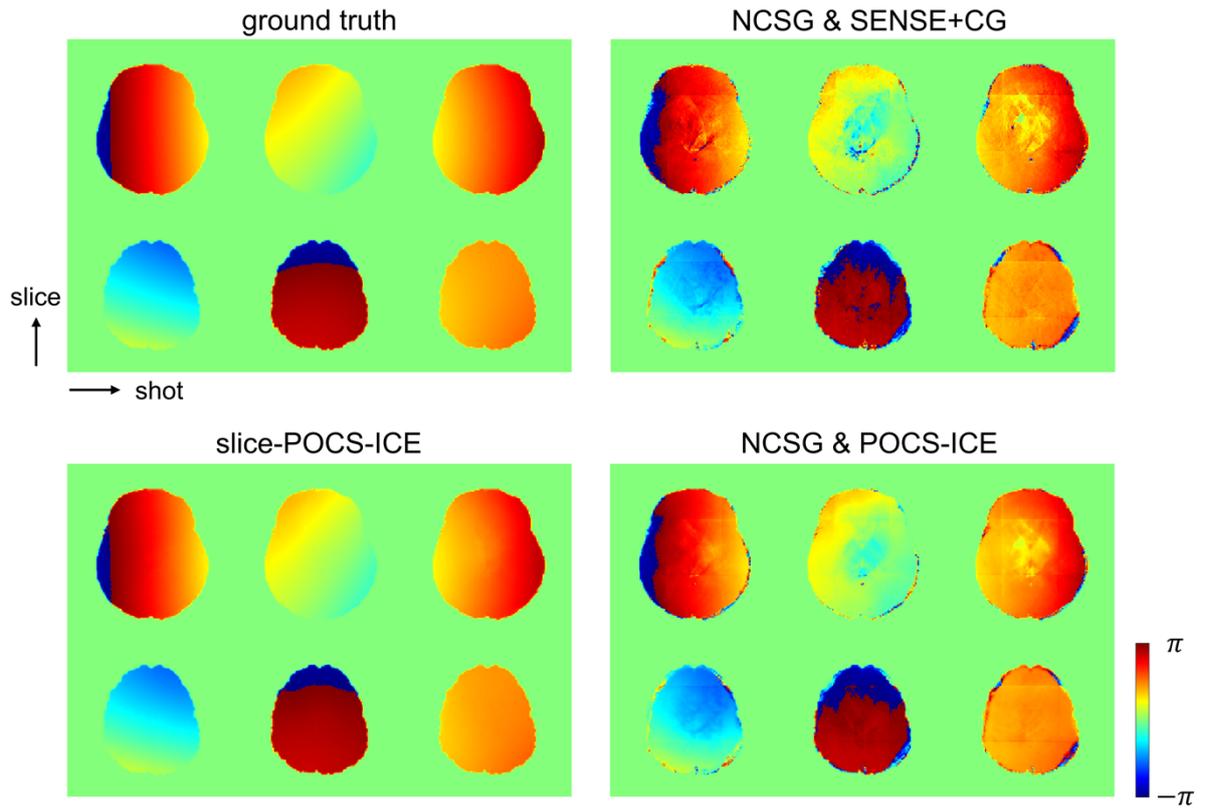

**Figure 4** Phase variation estimation results in numerical simulation for the [$N_{shot} = 3, N_{MB} = 2$] case. In each subfigure, the results of one method are shown, with different rows representing different slices, and different columns corresponding to different shots. The two selected slices are simultaneously excited in the simulation.

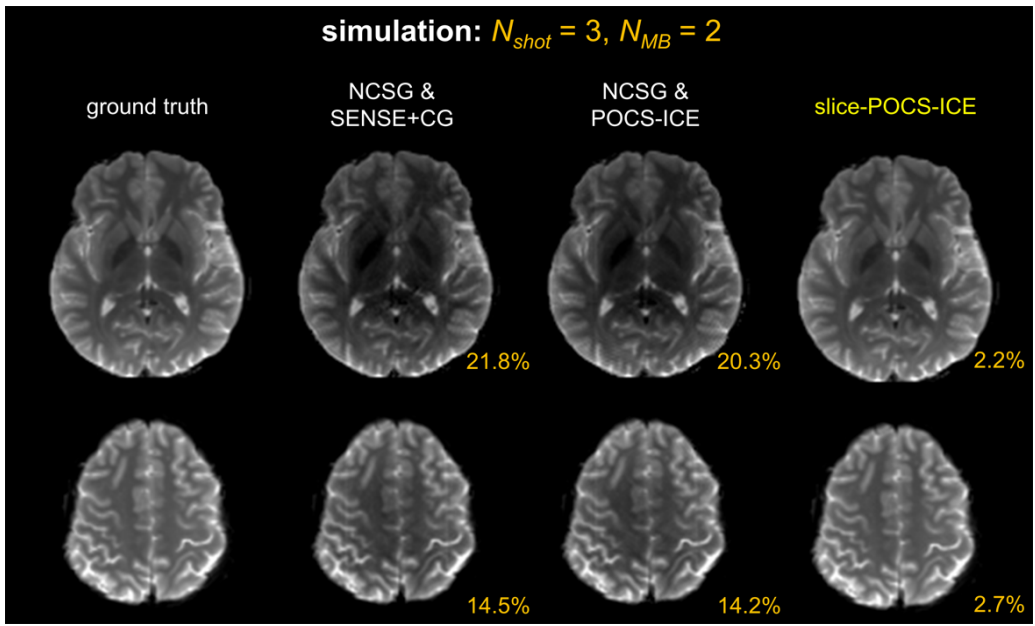

**Figure 5** Image reconstruction results in numerical simulation for the [$N_{shot} = 3, N_{MB} = 2$] case. The results of different methods are shown from left to right, with two simultaneously excited slices selected for display. The nRMSE values between the results and the ground truth are labeled.

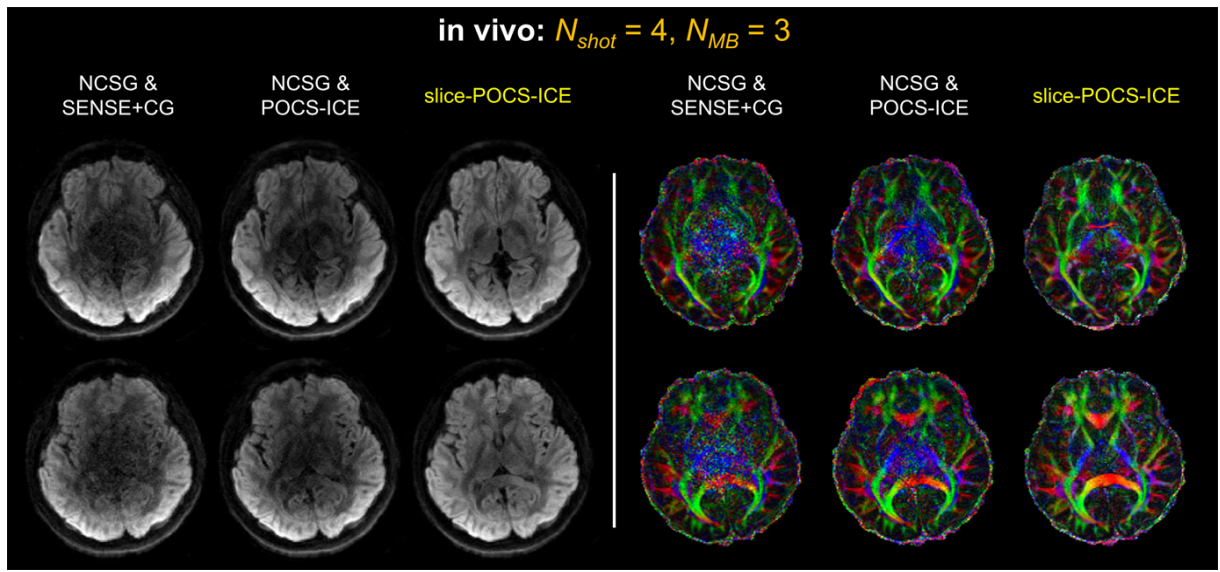

**Figure 6** Image reconstruction results of different methods in the first in vivo experiment for the [$N_{shot} = 4, N_{MB} = 3$] case. The left part displays the DWI images, and the right part displays the cFA images. For each method, two representative slices are shown.

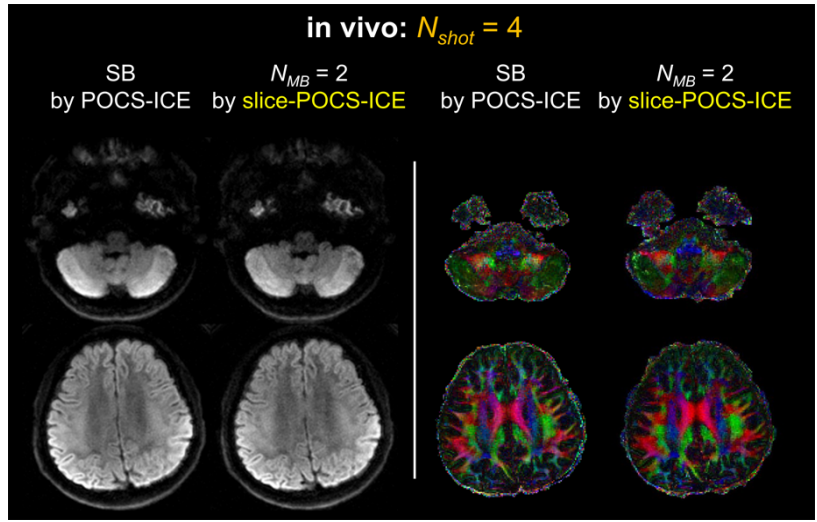

**Figure 7** Comparison of MB and SB reconstruction results in the first in vivo experiment. The DWI and cFA images are shown in the left and right panels, respectively. In each panel, the SB images were acquired using four shots and reconstructed by POCS-ICE, while the MB images were obtained from the [$N_{shot}$ = 4, $N_{MB}$ = 2] acquisition and reconstructed by slice-POCS-ICE. In the MB case, the two chosen slices were acquired simultaneously, whereas in the SB case, they were acquired separately.

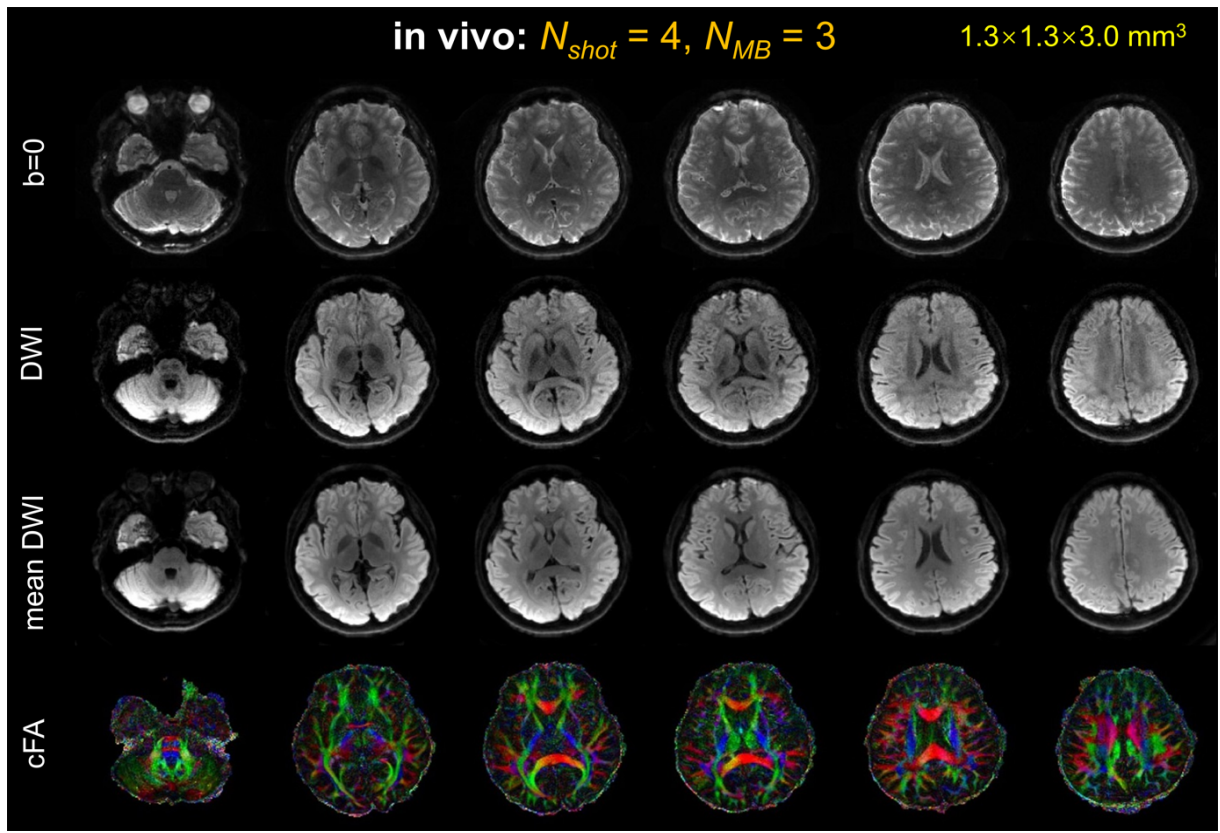

**Figure 8** Reconstruction results of slice-POCS-ICE in the first in vivo experiment for the $[N_{shot} = 4, N_{MB} = 3]$ case. Six slices from the skull base to the top are shown from left to right, and their b = 0, DWI, mean DWI, and cFA images are displayed.

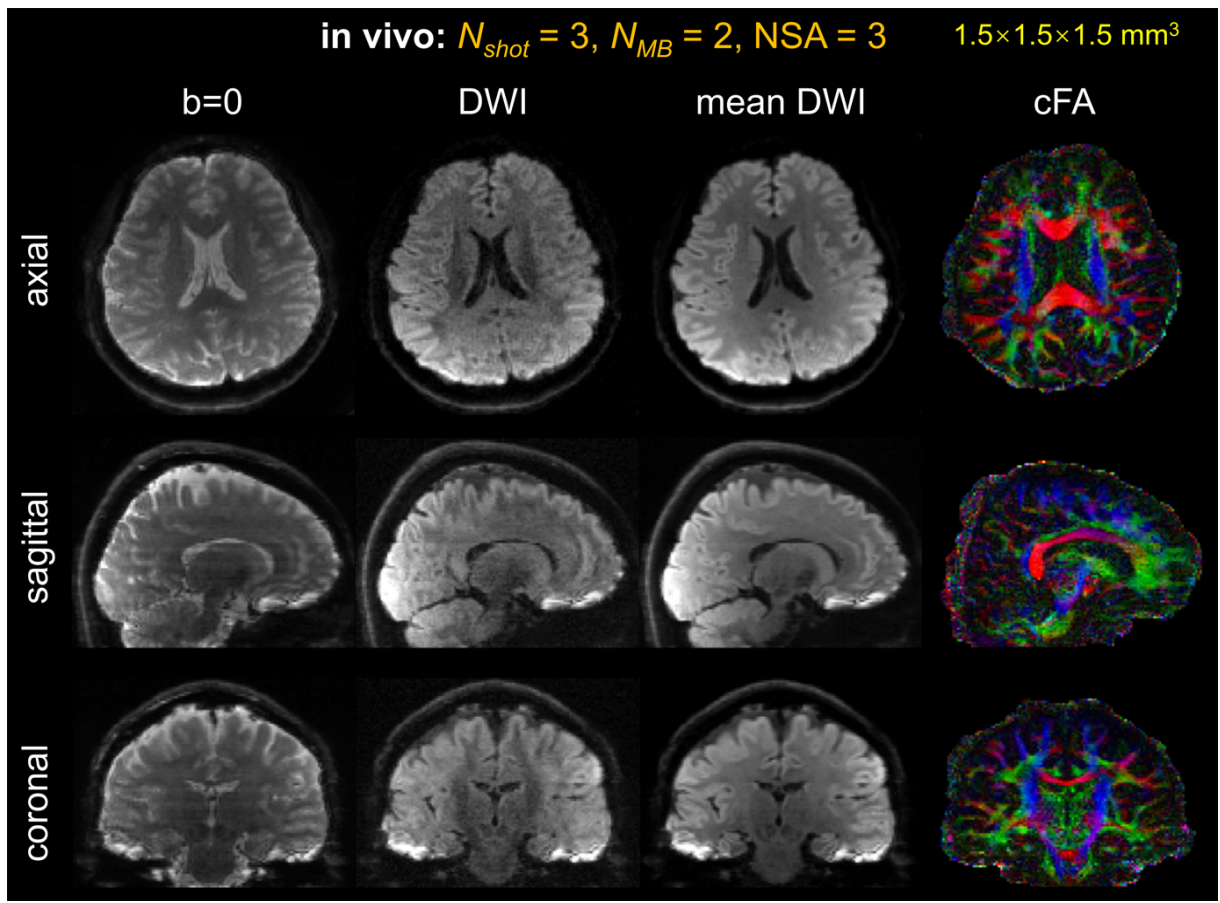

**Figure 9** Whole-brain images with isotropic 1.5 mm resolution from the third in vivo experiment. The images were obtained from the [$N_{shot}$ = 3, $N_{MB}$ = 2] acquisition and reconstructed by slice-POCS-ICE. One slice from each of the axial, sagittal, and coronal views is selected for display. The b = 0, DWI, mean DWI, and cFA images for each slice are shown.

**Table 1** Scanning parameters of the in vivo experiments.

Abbreviations: Exp., experiment; $N_{shot}$, number of shots; $N_{MB}$, number of simultaneously excited slices; $N_{diff}$, number of DWI scans; $N_{b0}$, number of b = 0 scans; $N_{slice}$, number of acquired slices; NSA, number of signal averages.

| Exp. | $N_{shot}$ | $N_{MB}$ | Resolution (mm³) | Sampling matrix | TE (ms) | TR (ms) | b value (s/mm²) | $N_{diff}/N_{b0}$ | $N_{slice}$ | Total scan time (min:sec) |
|---|---|---|---|---|---|---|---|---|---|---|
| 1 | 3 | 1 | 1.5×1.5×3.0 | 140×140 | 60 | 6000 | 800 | 12/1 | 46 | 3:54 |
|   | 3 | 2 | 1.5×1.5×3.0 | 140×140 | 60 | 3000 | 800 | 12/1 | 46 | 1:57 |
|   | 4 | 1 | 1.3×1.3×3.0 | 164×164 | 60 | 6000 | 800 | 10/1 | 46 | 4:24 |
|   | 4 | 2 | 1.3×1.3×3.0 | 164×164 | 60 | 3000 | 800 | 10/1 | 46 | 2:12 |
|   | 4 | 1 | 1.3×1.3×3.0 | 164×164 | 60 | 6000 | 800 | 10/1 | 51 | 6:36 |
|   | 4 | 3 | 1.3×1.3×3.0 | 164×164 | 60 | 3000 | 800 | 10/1 | 51 | 2:12 |
| 2 | 6 | 2 | 1.0×1.0×3.0 | 212×212 | 60 | 3000 | 800 | 10/1 | 46 | 3:18 |
| 3 | 3 | 2 | 1.5×1.5×1.5 | 140×140 | 60 | 6000 | 800 | 10/1 | 92 | 9:54 (NSA = 3) |

# Supporting Information

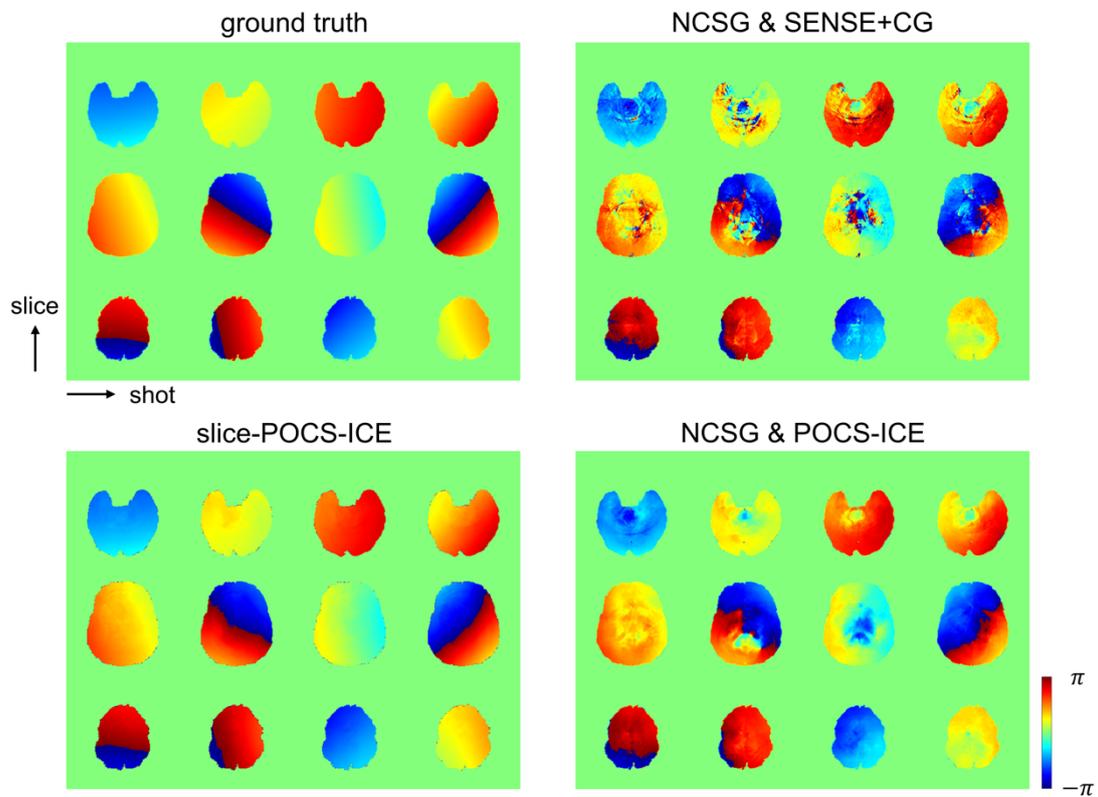

**Supporting Information Figure 1** Phase variation estimation results in numerical simulation for the [$N_{shot} = 4, N_{MB} = 3$] case. In each subfigure, the results of one method are shown, with different rows representing different slices, and different columns corresponding to different shots. The three slices selected are simultaneously excited in the simulation.

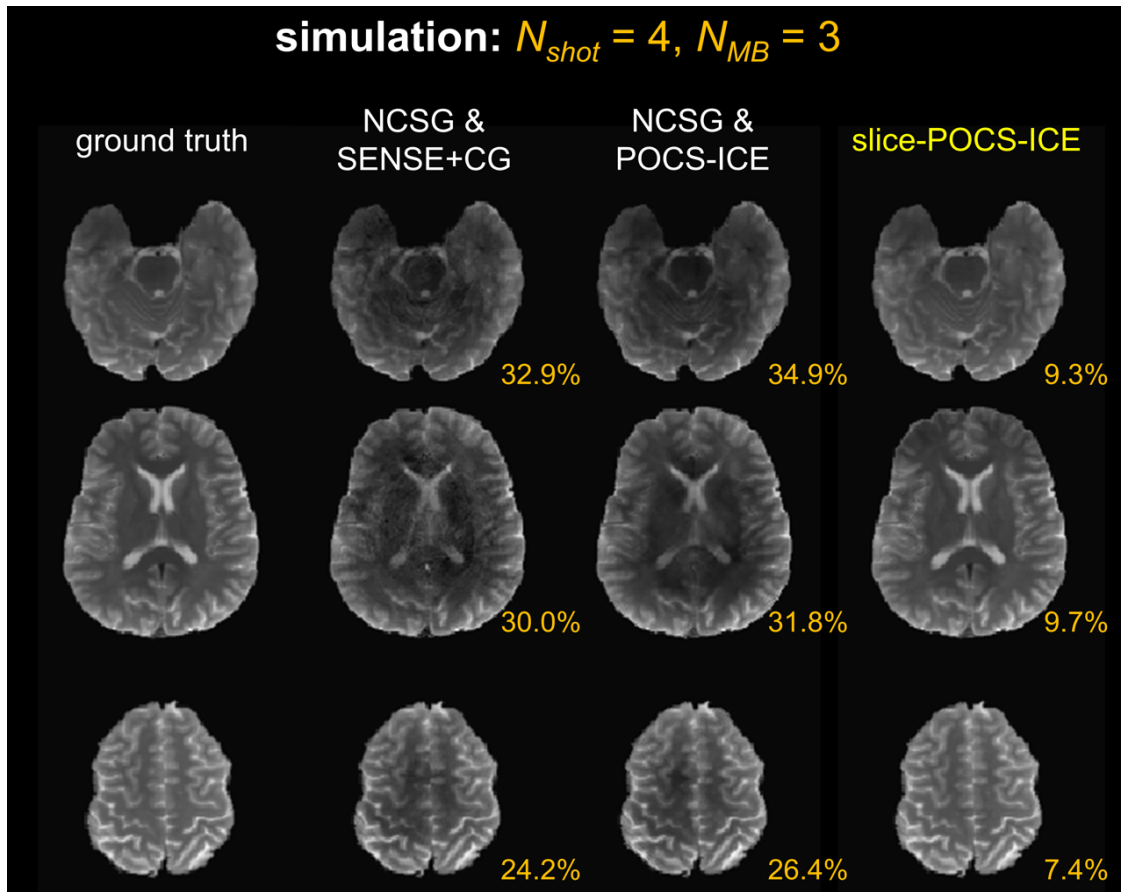

**Supporting Information Figure 2** Image reconstruction results in numerical simulation for the $[N_{shot}=4, N_{MB}=3]$ case. The results of different methods are shown from left to right, with three simultaneously excited slices selected for display. The nRMSE values between the results and the ground truth are labeled.

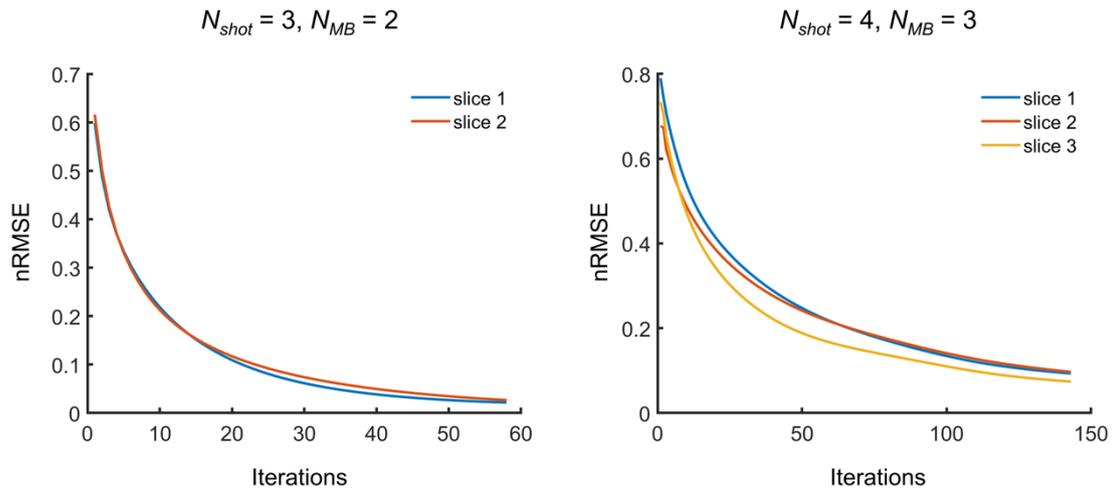

**Supporting Information Figure 3** Convergence performance of slice-POCS-ICE. The left and right panels show the $[N_{shot} = 3, N_{MB} = 2]$ and $[N_{shot} = 4, N_{MB} = 3]$ cases, respectively. In each panel, the nRMSE value of each slice with respect to the iteration numbers is plotted. The nRMSE values of different slices are distinguished by different colors.

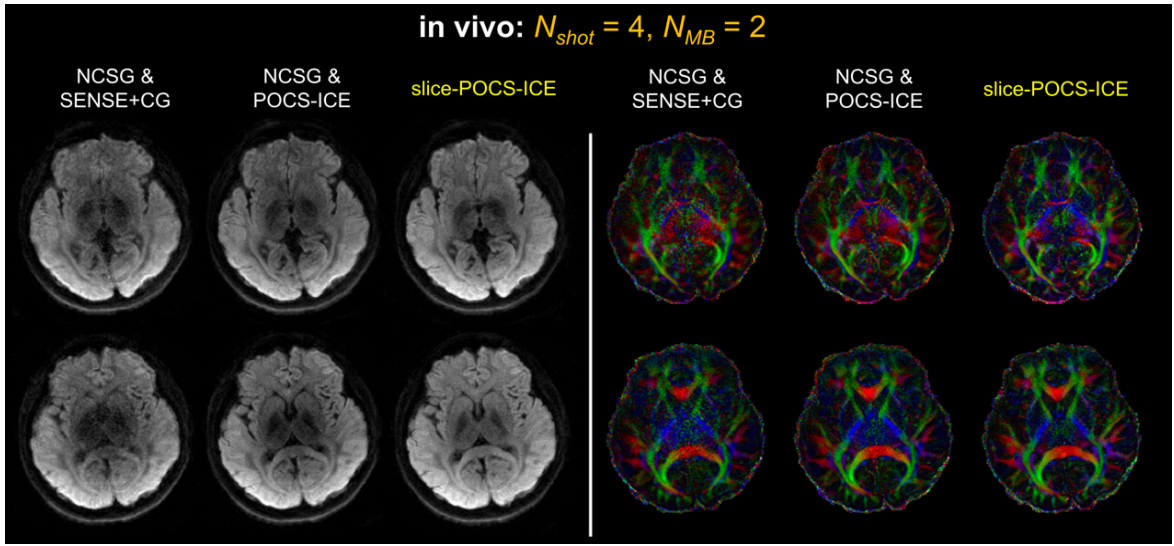

**Supporting Information Figure 4** Image reconstruction results of different methods in the first in vivo experiment for the $[N_{shot}=4, N_{MB}=2]$ case. The left part displays the DWI images, and the right part displays the cFA images. For each method, two representative slices are shown.

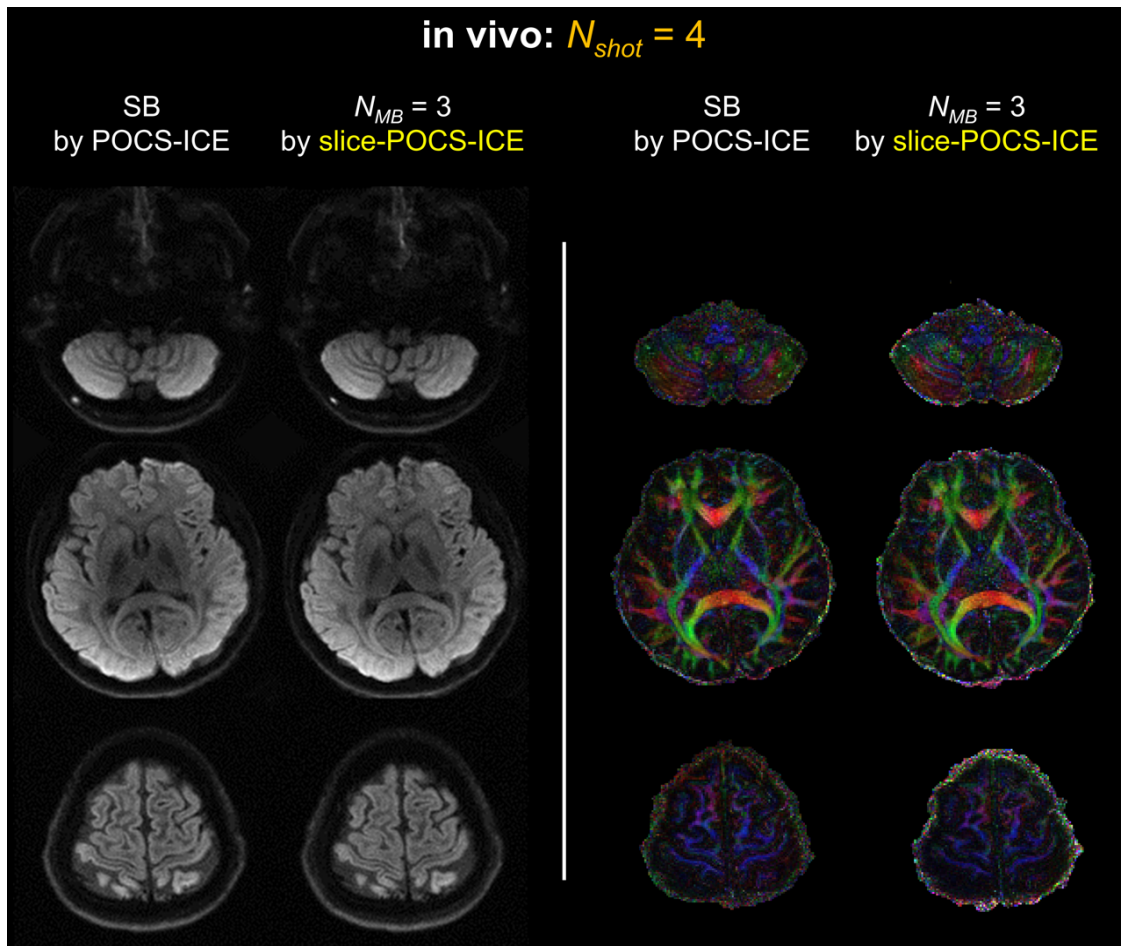

**Supporting Information Figure 5** Comparison of MB and SB reconstruction results in the first in vivo experiment. The DWI and cFA images are shown in the left and right panels, respectively. In each panel, the SB images were acquired using four shots and reconstructed by POCS-ICE, while the MB images were obtained from the [$N_{shot}$ = 4, $N_{MB}$ = 3] acquisition and reconstructed by slice-POCS-ICE. In the MB case, the three chosen slices were acquired simultaneously, whereas in the SB case, they were acquired separately.

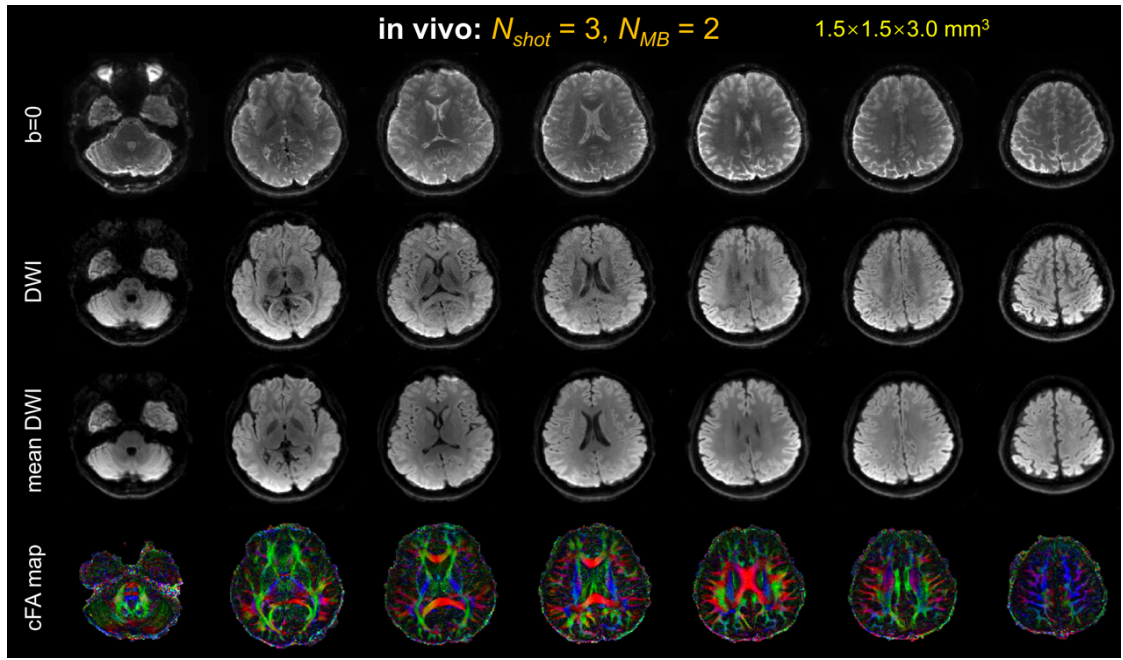

**Supporting Information Figure 6** Reconstruction results of slice-POCS-ICE in the first in vivo experiment for the [$N_{shot} = 3, N_{MB} = 2$] case. Seven slices from the skull base to the top are shown from left to right, and their b = 0, DWI, mean DWI, and cFA images are displayed.

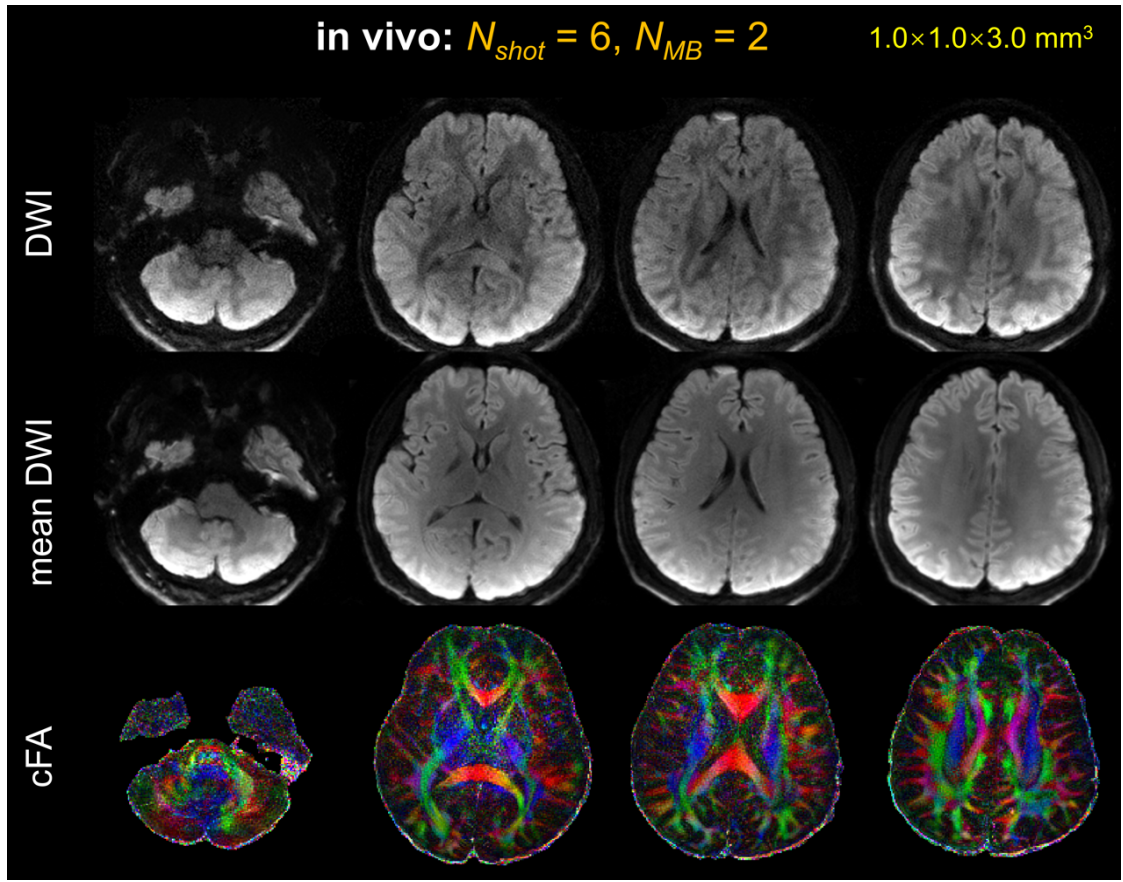

**Supporting Information Figure 7** High-resolution diffusion images with a resolution of 1×1×3 mm³ from the second in vivo experiment. The images were obtained from the [$N_{shot}$ = 6, $N_{MB}$ = 2] acquisition and reconstructed by slice-POCS-ICE. Four representative slices are shown from left to right, and their DWI, mean DWI, and cFA images are displayed.